\documentclass[preprint,aps,showpacs]{revtex4}
\usepackage{graphicx,amsfonts,amsmath}
\usepackage{epsfig}

\begin{document}

\title{\Large{Neutrino  Mass Through Concomitant Breakdown of the U(1) Chiral and Scale Symmetries  }}

\vspace{1.6 cm}

\author{Alex Gomes Dias\footnote{e-mail: alexdias@fma.if.usp.br}}
\affiliation{Instituto de F\'\i sica, Universidade de S\~ao Paulo, \\
C. P. 66.318, 05315-970 S\~ao Paulo, SP, Brazil. 
\vspace{1.6 cm}}

\date{\today}
\begin{abstract}
The possibility of generating neutrino mass through see-saw  mechanism involving U(1) chiral Peccei-Quinn and scale symmetries breakdown is discussed. We consider a generic scale invariant model which has three Majorana fermions and a complex scalar singlet, which might be the one responsible for an invisible axion,  and we perform a summation of all leading logarithmic radiative corrections to the tree level potential.  The effective potential so obtained is stable and drives the scalar field to a nonzero vacuum expectation value according to the Coleman-Weinberg mechanism. As a result, right-handed neutrinos gain mass at the Peccei-Quinn scale which is suggestive for explaining very light active neutrinos. We illustrate the whole idea with the addition of the Standard Model, and also a SU(3)$_L\otimes$U(1)$_X$ model in which the subgroup SU(2)$_L\otimes$U(1)$_Y$ is constrained to be broken as an effect of the effective potential. This last model presents electric charge quantization as well. 

\end{abstract}

\maketitle

\section{Introduction} 
\label{sec:intro}

The problem of neutrino's mass and mixing is now clearly seen as a central issue in model building in particle physics. This happened thanks to the enormous effort of the experimentalists during decades, to determine some of the neutrino's properties. The results that came out from the experiments point to very small masses, for explaining the oscillation phenomena, to the active neutrinos (those participating directly in electroweak interactions) and they have given more concrete reasons for considering some Standard Model (SM) extensions. We are led to think  of associating the eV scale, or less than that, to that particle masses according to the data obtained from solar and atmospheric experiments \cite{pdgneut}.

Mere introduction of right-handed components for the neutrinos in order to make them more similar to all the other known elementary particles as they appear in the SM is not satisfactory at all. This is because the scale $v_W\simeq 246$ GeV, at which the SM SU(2)$_L\otimes$U(1)$_Y$ symmetry breaks down to the electromagnetic factor, is clearly not suitable to be linked to particles with masses of eV order.  It is a mystery why an active neutrino belonging to the same group structure as the left-handed electron is so light in respect of it. A mechanism generating small masses to the active neutrinos seems, then, different from what we are used to think for most of the SM fields. It should contain more structure and should be associated, in principle, with an energy scale which can be much higher than $v_W$.

It is remarkable for neutrino that the probable nature of being a Majorana particle it should be related to its very small mass. When new physics at high energy scale is assumed we have a nonrenormalizable effective operator formed by the Higgs ($\Phi$) and the lepton ($L$) SU(2)$_L$ doublets as $ \Phi \Phi L L $ which is suppressed by a novel energy scale M \cite{sw79}. This operator breaks the lepton number and it gives mass to the active neutrinos of order $m_\nu\simeq v_W^2/M$ implying $M>>v_W$ for $m_\nu$ around eV scale or less than that. A way of realizing this has become known as see-saw mechanism which assumes the existence of  Majorana spinors and a new  physics at the very high energy $M$ \cite{seesaw}. The mechanism was first observed to be naturally contained in a SO(10) Grand Unified Theory (GUT) which does not preserve lepton number and it has, besides the SM particle content, right-handed neutrinos completing the representations. SO(10) could be  broken at scale $M$, generating a low energy effective theory which is the SM model plus heavy fields with  right-handed neutrinos among them. These right-handed neutrinos couple with the singlet formed by the Higgs and lepton SU(2)$_L$ doublets. Such mixing between right and left-handed neutrinos belonging to different scales would then be the explanation for the tiny masses of the active neutrinos. This is the well known type I see-saw mechanism and it is a general outcome of theories containing heavy  right-handed neutrinos mixing with the active  left-handed neutrinos.

An interesting high energy scale that may be connected with the neutrino's mass is the one associated with the invisible axion \cite{invaxion}, which is the pseudo Goldstone boson resulting from the  spontaneous breakdown of the  global anomalous U(1)$_{PQ}$ Peccei-Quinn symmetry \cite{pq,axion}. Such symmetry is admitted to exist for solving the strong CP problem. From astrophysical and cosmological constraints on the axion, the scale $v_{PQ}$ at which U(1)$_{PQ}$ is broken is restricted to be in the range $10^{9}$ GeV $\leq v_{PQ} \leq 10^{12}$ GeV \cite{pdgaxion}. In the simplest implementation of the idea, $v_{PQ}$ arises from the vacuum expectation value (VEV) of a complex scalar field $\phi$, without direct couplings with matter, carrying a charge of U(1)$_{PQ}$. Stabilization of the strong CP problem solution through an axion is also needed, because the whole mechanism is very sensitive to the impact of effective nonrenormalizable operators, suppressed by the Planck scale, $M_{Pl}\simeq 10^{19}$ GeV, which must arise from quantum gravity. One way out of this is to postulate gauge discrete $Z_N$ symmetries \cite{kw} giving some protection for the global factor U(1)$_{PQ}$ against gravity \cite{znaxion,331ax1}.  

If  according to some symmetry like, for example, the $Z_N$ symmetry, and consequently the U(1)$_{PQ}$, the complex scalar field $\phi$ may couple with right-handed neutrinos at the tree level, then their masses  would be generated at the scale $v_{PQ}$ and the see-saw mechanism should occur. In the absence of such tree level masses, higher dimensional operators ought to generate the leading mass terms combining the scales $v_{PQ}$, M and $M_{Pl}$ \cite{nusaxion}.

Concerning the breakdown of the U(1)$_{PQ}$, which is assumed to happen through the usual way; when a tree level potential $V(\phi)$ has a nontrivial minimum at $\langle \phi \rangle \approx v_{PQ}$ due to the presence of a bilinear term like $\mu^2\phi^\dagger\phi$ whose signal is ad hoc chosen to produce the desired shape for $V(\phi)$. The introduction of the mass square parameter $\mu^2$ is the additional cost, besides the eventual imposition of new symmetries, which in some sense represents what we call "trading one problem for another" once the degree of indetermination of the theory is increased. Then, a more satisfactory way of realizing spontaneous symmetry breakdown (SSB) is required. In fact, such a way can be realized according to the general Coleman-Weinberg mechanism (CWM) which puts radiative corrections as the origin of SSB \cite{colwein}. The CWM is based on the observation that the tree level conformal symmetry of scale invariance, which forbids parameters with mass dimension in the Lagrangian, is  violated by quantum corrections and that when some conditions are satisfied SSB happens by means of an effective potential.

In this work we use the idea of CWM with a summation over all leading-logarithm contributions in a generic sector of a scale invariant theory containing Majorana fermions and a scalar complex field. The summation of  leading-logarithms  results in an improved calculation of the effective potential because it extracts all available information contained in the renormalization group equations  \cite{eliasprl,eliasnpb,elias05}. And so a more accurate answer for the question of the mechanism of radiative corrections in the process of SSB can be obtained.

Here the scalar field effective potential will be determined with the conditions which must be satisfied in order to justify the calculation inside the perturbative regime. We vary the scalar couplings with the Majorana fermions from small values to the ones near the perturbative limit in order to know if SSB can happen or not through CWM. As we will see the answer turns out to be positive for all the set of parameters tested so that, even for the three heaviest possible right-handed neutrinos, CWM is operative in explaining mass generation for these particles. As a consequence SSB of U(1)$_{PQ}$ is mandatory and we have a solution for the strong CP problem together with the generation of very light active neutrinos by means of see-saw mechanism in an economic way in one model. For the several values of the right-handed neutrino masses we have a prediction for the heavy scalar axion-partner. In some cases this particle cannot decay into right-handed neutrino and it can bring some interest for cosmological studies. We perform the study  considering a generic model, that is the Standard Model plus three Majorana fermions and a complex singlet, and also a SU(3)$_L\otimes$U(1)$_X$ model in which electric charge quantization occurs as well.

The work is organized as follows. In Sec. \ref{sec2} the generic model is presented and some of its main points are discussed; Sec. \ref{sec3}  deals with the calculation of the effective potential with the  summation of leading-logarithmic contributions using a general form of the relevant beta functions. We also study some possible set of parameters to the  effective potential and the resulting particle masses; Sec. \ref{sec4} is an application to an electroweak SU(3)$_L\otimes$U(1)$_X$ model and discussion of results are presented in Sec. \ref{sec:disc}.

\section{The generic model}
\label{sec2}

We present here a model that can be considered as a sector of some more specific theory, so that the results we have achieved are general in this sense. The relevant interaction Lagrangian is
\begin{equation}
{\cal{L}}= \sum_{\alpha,\beta=1}^3\frac{ \lambda^{\prime}_{_{ \alpha\beta}}}{2}\left[\phi^* \overline{( N^{\prime}_{_{ \alpha R}})^c} N^{\prime}_{_{ \beta R}} + \phi \overline{( N^{\prime}_{_{ \alpha L}})^c} N^{\prime}_{_{ \beta L}} \right]+ \lambda_\phi (\phi^*\phi)^2
\label{nuax}
\end{equation}
Here $N^{\prime}_{_{ \alpha }}$ are the Majorana spinors from where right-handed neutrinos are composed, $\phi$ is a complex scalar field whose imaginary part will be the major part of the axion field and $\lambda^{\prime}_{_{ \alpha\beta}}$ is a symmetric matrix which we assume to be real to simplify the analysis.

In principle, we could have an arbitrary number of $N^{\prime}_{_{ \alpha }}$. Keeping in mind the idea of the see-saw mechanism we work with the minimal scenario, where  for each light active neutrino there must be a heavy right-handed neutrino. Although, at the present, it is said that the neutrino oscillation experimental data are still compatible with one active neutrino having null mass  we adopt here the point of view that this is not the case. 

There is no reason for the matrix $\lambda^{\prime}_{_{ \alpha\beta}}$ to be diagonal. But we can make it diagonal rotating  the $N^{\prime}_{_{ \alpha }}$ and defining a new ${\bf \lambda }$ through
\begin{equation}
\lambda = U^T_N \lambda^{\prime}U_N 
\label{ld}
\end{equation}
So that  Eq. (\ref{nuax})  becomes 
\begin{equation}
{\cal{L}}= \sum_{\alpha=1}^3\frac{ \lambda{_{ \alpha}}}{2}\left[\phi^* \overline{( N_{_{ \alpha R}})^c} N_{_{ \alpha R}} + \phi \overline{( N_{_{ \alpha L}})^c} N_{_{ \alpha L}} \right]+ \lambda (\phi^*\phi)^2
\label{nuaxd}
\end{equation}
Although the parameters of this rotation do not appear in this sector anymore they are transfered to other sectors, like the one of  the Yukawa interaction needed for the see-saw mechanism, in which we are interested. Then, there are interactions among the SM lepton doublets $L_{La}$, the Higgs doublet $\Phi$ and the $N_{_{ \alpha }}$ according to the term
\begin{equation}
{\cal{L}}^{Dirac}= G^N_{a \alpha}(U_N)_{\alpha\beta} \overline{L_a{_L}} \tilde{\Phi} N_{\beta _{R}} 
\label{nuNdirac}
\end{equation}
where $\tilde{\Phi}\equiv\epsilon\Phi$ and we have made the explicit appearance of the matrix $U_N$. After the condensation of the Higgs field, $\langle \Phi\rangle_2= v_{_W}/\sqrt2$, SSB of the electroweak symmetries Eq. (\ref{nuNdirac}) yields the Dirac mass matrix 
\begin{equation}
{\cal{M}}^{D}_{a\beta}=  G^N_{a\alpha}[U_N]_{\alpha\beta}  \frac{v_{_W}}{\sqrt2}
\label{mdirac}
\end{equation}
For our purposes, in the renormalization group equations, it is more convenient to work with the Lagrangian as in Eq. (\ref{nuaxd}) since the relevant beta functions are also diagonal in this case. But we have to keep in mind that the effect of the rotation must be carried out in the Dirac type  mass matrix as in Eq. (\ref{mdirac}). 

Finally, we mention that ${\cal{L}}$ above has the following global symmetry which we identify as the U(1)$_{PQ}$ 
\begin{equation}
 N^{\prime} \rightarrow e^{i\delta\gamma_5} N^{\prime}, \hspace{1.0 cm}  \phi \rightarrow e^{2i\delta} \phi
\label{upq}
\end{equation}
Now we begin the calculation of the effective potential which we are supposing to break this and the scale symmetry.

\section{Summation of the leading-logarithms for the effective potential}
\label{sec3}

The basic points for computing the effective potential with summation of  leading-logarithmic at one loop was given in Refs. \cite{eliasprl,eliasnpb}, where the case of the Higgs potential of the SM and the scalar electrodynamics were carried out in detail. We will follow the same steps here aiming the model in Eq. (\ref{nuaxd}) which has not yet been considered from the CWM point of view at the present literature, thus justifying the review of the detailed calculation.  

Defining the complex scalar field in Eq. (\ref{nuaxd}) as $\phi(x)=\varphi(x) e^{i\frac{a(x)}{v}}/\sqrt2$, with both $\varphi(x)$ and $a(x)$ real and $v$ the scale where $\phi$ is supposed to condense, we have that the tree level scale invariant potential in Eq. (\ref{nuaxd}) is, then,
\begin{equation}
V=\frac{\lambda_\phi}{4} \varphi^4 
\label{vtree}
\end{equation}
Radiative corrections are such that the effective potential in  any order of perturbation theory is a function like 
\begin{eqnarray}
V_{eff} & \approx & {\pi^2} \varphi^4 S(y, x_\alpha ,L) \nonumber\\ \nonumber\\ 
& = & {\pi^2} \varphi^4 [ y + {\cal{O}}(y^2, x_\alpha^2, yx_\alpha^2,L)]
\label{vef}
\end{eqnarray}
where for convenience we have written:
\begin{eqnarray}
& & y = \frac{\lambda_\phi}{4\pi^2}; \hspace{.6 cm} 
x_\alpha = \frac{\lambda_\alpha}{4\pi^2}, \hspace{.3 cm}\alpha=1,2,3; \hspace{.6 cm} \nonumber\\
\nonumber\\
& & L=\ln\left(\frac{\varphi^2}{\mu^2}\right) 
\label{xyL}
\end{eqnarray}
The  $S(y, x_\alpha ,L)$ represents an infinite series containing corrections due to the fields $N_\alpha$ and $\phi$ only. Of course, the exact effective potential would have a function $S$ involving  all other coupling constants in the theory. Their contributions start to appear at two loop order and they came in powers combined with $x_\alpha $ and $y$. We must say that we have made at this point the assumption that all other coupling constants are subdominant compared to $\lambda_\alpha$ and $\lambda_\phi$ so that the $V_{eff}$ is well described by Eq. (\ref{vef}). Now, we do what is called the leading-logarithm approximation which consists in considering $ S(y, x_\alpha ,L)\approx  S_{LL}(y, x_\alpha ,L)$, with the subseries $S_{LL}(y, x_\alpha ,L)$ having terms only in $L^n$ whose sum of aggregate powers of $x_\alpha $ and $y$ is equal to $n+1$, with obviously $n\geq 0$. Then we write a sum for the  leader logarithmic terms as

\begin{eqnarray}
S_{LL}(y, x_\alpha ,L) & = & \sum_{k,l,m,n=0}^\infty C_{klmn}  x^k_1x^l_2 x^m_3 y^n L^{k+l+m+n-1}, \hspace{.6 cm} (C_{0000}=0) 
\nonumber\\ \nonumber\\ 
& = &  yS_0 + \sum_{\underset{(k+l+m\geq 1)}{k,l,m}}  x^k_1x^l_2x^m_3  L^{k+l+m-1} S_{klm}  
\label{sll}
\end{eqnarray}
Where the partial sums are defined 
\begin{eqnarray}
& & S_0 = \sum_{n=1}^\infty  C_{000n} u^{n-1} 
\label{s0} \\ \nonumber\\ 
& & S_{klm} = \sum_{n=0}^\infty C_{klmn} u^{n} 
\label{s0sklm}
\end{eqnarray}
with $u=yL$.
The invariance of the effective potential with renormalization scale $\mu$, i. e. $ d\, V_{eff}^{LL}/ d\mu=0$, leads to the following equation in the leader logarithmic approximation 

\begin{eqnarray}
\Big{[} -2 \frac{\partial}{\partial L} + \beta_{ y}\frac{\partial}{ \partial y}  + \sum_{x_\alpha=1}^3 \beta_{ x_\alpha}\frac{\partial}{\partial  x_\alpha}  -4 \gamma_\phi  \Big {]} S_{LL} (y, x_\alpha, L) =0,   
\label{rgsll}
\end{eqnarray}

In the one loop approximation the beta function $\beta_x$ and the anomalous dimension $\gamma_\phi$ of the scalar field have the following generic form  
\begin{eqnarray}
& & \beta_{x_\alpha}= \frac{\mu}{4\pi^2}\frac{d\, \lambda_\alpha^2}{d\, \mu}  = f x_\alpha \sum_{\beta\not=\alpha} x_\beta  
\label{bgx} \\ \nonumber\\ 
& & \gamma_\phi=\frac{\mu}{2Z_\phi}\frac{dZ_\phi}{d\mu} = e\sum_{\alpha=1}^3 x_\alpha  
\label{bgg}
\end{eqnarray}
The coefficients $f$ and $e$ are determined through computation of Feynman diagrams; $Z_\phi$ is the wave function renormalization of the scalar field $\phi$. Observe that just for one Majorana fermion the beta function in Eq. (\ref{bgx}) is zero, since there is no $\beta\not=\alpha$. This is because in this case, like in supersymmetry, there is the same number of bosonic and fermionic degrees of freedom.   

We can extract $\beta_{y}$ from the renormalized one loop effective potential which has the form,

\begin{equation}
V_{eff}^{1loop}= \pi^2 \varphi^4  \left[ y + \frac{1}{2} \left(a y^2+c\sum_{\alpha=1}^3 x_\alpha^2 \right) L \right],
\label{v1l} 
\end{equation}
$a$ and $c$ can be computed directly from the general expression for the one loop corrections for the potential. Invariance of the effective potential under changing the renormalization scale, i. e. $V_{eff}^{1loop}(\mu)= V_{eff}^{1loop}(\mu+\delta\mu)$, permits to deduce that  
\begin{eqnarray}
\beta_{y}  & = & \frac{\mu}{4\pi^2}\frac{d\, \lambda_\phi}{d\, \mu}  
= ay^2 +c\sum_{\alpha=1}^3 x_\alpha^2 +4y \gamma_\phi  \nonumber \\\nonumber \\ 
& = & ay^2 + c\sum_{\alpha=1}^3 x_\alpha^2 +4ey
\sum_{\alpha=1}^3 x_\alpha 
\label{bgy} 
\end{eqnarray}

Now substituting $S_{LL}$ as Eq. (\ref{sll}) in Eq. (\ref{rgsll}) and using Eqs. (\ref{bgx}), (\ref{bgg}), (\ref{bgy}) we can obtain relations among the coefficients $C_{klmn}$. These relations can be writen as differential equations for  $S_0$ and $S_{klm}$ which can be solved recursively. Before doing this explicitly we see that comparing Eqs. (\ref{v1l}), (\ref{sll2}) and from interchanging symmetry on $x_\alpha$ in the series we have  
\begin{eqnarray}
&  & C_{0001} =  1; \,\,\,\,\,\,\, C_{1000} = C_{0100} = C_{0010} = 0  \nonumber \\ 
&  & C_{1100} = C_{10100} = C_{0110} = 0;  \,\,\,\,\,\,\, C_{0002} = \frac{a}{2}; \nonumber \\ 
&  & C_{2000} = C_{0200} = C_{0020} = \frac{c}{2}  
\label{cs} 
\end{eqnarray}
Now from Eq. (\ref{rgsll}) each term in powers of $x_\alpha$, $y$ and  $L$ must have a null coefficient.

$1.)$ from terms like $y^nL^{n-2}$ 
\begin{eqnarray}
&  & 2C_{000n} - a C_{000n-1}=0;  \,\,\,\,\,\,\, n\geq 2 
\label{c000n} 
\end{eqnarray}
and with $C_{0001}$ in (\ref{cs}) we find that 
\begin{eqnarray}
C_{000n} =  \left(\frac{a}{2} \right)^{n-1}, \,\,\,\,\,\,\, n\geq 1 
\label{c000nf} 
\end{eqnarray}

$2.)$  terms like $x_1y^nL^{n-1}$ 
\begin{eqnarray}
- 2 n C_{100n}  + a(n-1)C_{100n-1}  + 4e(n-1)C_{000n}=0, \,\,\,\,\, n\geq 1
\label{c100n}
\end{eqnarray}

$3.)$  terms like $x_1^2y^nL^{n}$ 
\begin{eqnarray}
- 2 (n+1) C_{200n} + c(n+1)C_{000n+1}   + a(n-1)C_{200n-1}  + 4e(n-1)C_{100n}=0, \,\,\, n\geq 1
\label{c200n}
\end{eqnarray}

$4.)$  terms like $x_1 x_2 y^nL^{n}$ 
\begin{eqnarray}
- 2 (n+1) C_{110n}  + a(n-1)C_{110n-1}  + 2\left(4e(n-1)+f \right)C_{100n}=0, \,\,\, n\geq 1
\label{c110n}
\end{eqnarray}

The remaining relations involving all other coefficients can be obtained as well. But the expressions written in Eqs. (\ref{c000nf}), (\ref{c100n}), (\ref{c200n}) and (\ref{c110n}) are sufficient for the approximation we are going to work. For example, from the coefficients of terms like $x_1^k y^n L^{k+n-2}$ we have 
\begin{eqnarray}
- 2 (k-1) C_{k000}  + cC_{k-2001}  - 4eC_{k-1000}=0, \,\,\,\,\,k\geq 2, \,\, n\geq 1
\label{ck000}
\end{eqnarray}
and
\begin{eqnarray}
- 2 (k+n-1) C_{k00n}  + a(n-1)C_{k00n-1}  + 4e(n-1)C_{k-100n} + c(n+1)C_{k-200n-1} =0, \nonumber \\  k\geq 3, \,\, n\geq 1
\label{ck00n}
\end{eqnarray}

The first partial sum, $S_0$, can be obtained easily using the relation given by Eq. (\ref{c000nf}) in Eq. (\ref{s0}) which furnishes, defining $w=1-\frac{a}{2}u$ 

\begin{equation}
S_{0}  =\frac{1}{w}
\label{s0f}
\end{equation}

Multiplying Eq. (\ref{c100n}) by $u^{n-1}$ we get the following differential equation 

\begin{equation}
aw  \frac{dS_{100}}{dw} -4e(1-w) \frac{dS_0}{dw} = 0
\label{dS1}
\end{equation}
whose solution, with the boundary condition $S_{100}(u=0)=S_{100}(w=1)=C_{1000}=0$, is

\begin{equation}
S_{100}= \frac{2e}{a} \left[ \frac{1}{w^2} -\frac{2}{w} + 1 \right]
\label{S1}
\end{equation}

Multiplying Eq. (\ref{c200n}) by $u^{n}$ we get the equation 

\begin{equation}
2\left[ w(1-w) \frac{d}{dw} - 1 \right] S_{200} - c\left[ (1-w)\frac{d}{dw} - 1\right]S_0 - 4e \left[(1-w)\frac{d}{dw} + 1\right] S_{100} = 0 
\label{dS2}
\end{equation}
whose solution, with the boundary condition $S_{200}(w=1)=C_{2000}=\frac{c}{2}$, is

\begin{equation}
S_{200}= \frac{1}{6a} \left[ 12 e^2 \frac{1}{w^3} +  \left( ca - 28 e^2 \right) \frac{1}{w^2} +  \left( ca + 20 e^2 \right) \frac{1}{w} +  ca - 4 e^2 \right] 
\label{S2}
\end{equation}

Doing just the same with Eq. (\ref{c110n}) we have 

\begin{equation}
2\left[ w(1-w) \frac{d}{dw} - 1 \right] S_{110} - 2\left[ 4e(1-w)\frac{d}{dw} + 4e -f\right] S_{100} = 0 
\label{dS11}
\end{equation}
and the solution, with the boundary condition $S_{110}(w=1)=C_{1100}=0$, is

\begin{equation}
S_{110}= \frac{2}{3a} \left[ 6 e^2 \frac{1}{w^3} +  \left( ef - 14 e^2 \right) \frac{1}{w^2} +  \left( 10 e^2 - 2ef \right) \frac{1}{w} +  ef - 2 e^2 \right] 
\label{S11}
\end{equation}

The partial sums in Eqs. (\ref{s0f}), (\ref{S1}), (\ref{S2}) and (\ref{S11}) are sufficient for obtaining $S_{LL}$ till the quadratic order in $x_\alpha$ summing the first leading logarithms. By writing down the expression truncated up to this order, we observe that from the fact that the beta functions for $x_\alpha$ in Eq. (\ref{bgx}) have the same form and the partial sums $S_{klm}$ are totally symmetric, so that 

\begin{eqnarray}
S_{LL} \approx yS_0 + \sum_{\alpha}x_{\alpha} S_{100} + \sum_{\alpha} x_{\alpha} ^2 LS_{200} + \frac{1}{2}\sum_{\alpha}\sum_{\beta\not=\alpha} x_{\alpha}x_{\beta} LS_{110}  
\label{sll2}
\end{eqnarray}

Expanding Eq. (\ref{sll2}) in $L$ until terms of order $L^4$, which is the same as keeping in the expansion terms proportional at most to $y^5$, we have then 

\begin{eqnarray}
V_{eff}^{LL} & = & {\pi^2} \varphi^4 [S_{LL}(x_{\alpha}, y, L) + K ] 
\nonumber\\\nonumber\\
& \approx & {\pi^2} \varphi^4 [ y  + B L + C L^2 + D L^3 + E L^4 + K ] 
\label{vefll2}
\end{eqnarray}

We have added a counter term with the finite coefficient $K$, to be determined, in order to make easy the $V_{eff}^{LL}$  fourth-derivative renormalization condition  which defines the scalar field four-point self-interaction (see below); and we called the functions of the coupling constants 

\begin{equation}
B= \frac{a}{2}y^2 + \frac{c}{2} \sum_{\alpha} x_{\alpha}^2 
\label{B} 
\end{equation}
\begin{equation} 
C= \frac{a}{4}y  \left( ay^2 + 2ey\sum_{\alpha} x_{\alpha} + c\sum_{\alpha} x_{\alpha}^2   \right)
\label{C} 
\end{equation} 
\begin{equation} 
D= \frac{a}{24}y^2   \left( 3a^2y^2 + 12aey\sum_{\alpha} x_{\alpha} + 4(ac+2e^2)\sum_{\alpha} x_{\alpha} ^2 +  2(ef+4e^2) \sum_{\alpha}\sum_{\beta\not=\alpha}x_{\alpha}x_{\beta} \right)
\label{D} 
\end{equation}
\begin{equation} 
E= \frac{a^2}{48}y^3  \left( 3a^2y^2 + 18aey\sum_{\alpha} x_{\alpha} + 
(5ac+28e^2)\sum_{\alpha} x_{\alpha} ^2 + 4(ef+7e^2) \sum_{\alpha}\sum_{\beta\not=\alpha} x_{\alpha}x_{\beta} \right)
\label{E} 
\end{equation}

Two renormalization conditions are used to fix the value of $K$ in Eq. (\ref{vefll2}) and to determine $y$ as a function of the $x_{\alpha}$. They are 

\begin{equation} 
\frac{d^4\, V_{eff}^{LL}}{d\, \varphi^4} {\Big{\vert}}_{\varphi=\mu} = 24\pi^2y
\label{d4v} 
\end{equation}

\begin{equation} 
\frac{d\, V_{eff}^{LL}}{d\, \varphi}  {\Big{\vert}}_{\varphi=\mu} = 0
\label{dv} 
\end{equation}

So that, it is determined

\begin{equation}
K = -\frac{25}{6} B  - \frac{35}{3} C  - 20 D  - 16E   
\label{K}
\end{equation}

and 

\begin{equation}
y = \frac{11}{3} B  + \frac{35}{3} C  + 20 D  + 16E   
\label{y}
\end{equation}

Eq. (\ref{y}) is a polynomial equation of fifth degree. We must assure that among its solutions there is at least one inside the perturbative range, i. e., satisfying $\pi y < 1$, for some of the possible values of $x_{\alpha}$. For the interaction Lagrangian we have used here, the values of the numerical constants, as it can be shown, are: $a=5$, $c=-1/32$, $f=1/8$ and $e=1/16$. There are solutions of Eq. (\ref{y}) for $y$ inside the perturbative range for some choice of  $\lambda_\alpha$, as it can be seen in Table \ref{table1}. Just for simplicity, we have taken all $\lambda_\alpha$ being equal. In this case, we see that there is always a unique real positive solution for $y$.

\begin{table}
\begin{tabular}{||c|c|c|c|c||}\hline
$\pi x_\alpha$ & $y$ & $\pi y$ & $m_\varphi/\langle\varphi\rangle$ & $m_{N_\alpha}/\langle\varphi\rangle$\\
\hline
0.001 & 0.0650 & 0.204 & 0.980 &  0.039 \\ 
0.010 & 0.0649 & 0.204 & 0.980 &  0.120 \\
0.100 & 0.0643 & 0.202 & 0.968 &  0.396 \\ 
0.500 & 0.0656 & 0.206 & 0.923 &  0.886 \\ \hline
\end{tabular}
\caption{Values for $y$ and the relative masses  $m_\varphi/\langle\varphi\rangle$ and $m_{N_\alpha}/\langle\varphi\rangle$ for some choices of $\lambda^2_\alpha/4\pi= \pi x_\alpha $.}
\label{table1}
\end{table}

Although for the choices of  $x_\alpha$ here we have $\pi y\approx 0.2$, we see that the results show a relative strong coupling for the scalar $\phi$ self-interaction. The coupling constant $\lambda_\phi$ seems do not present a significant dependence on $x_\alpha$. This is unexpected. As we have discussed at the beginning, maintaining only the simple one loop approximation, heavy fermions were seen to put a bound on the effective potential stability. Also, we observe from the last line of Table \ref{table1} that it is possible to have a situation where the mass of the Majorana heavy fermions are such that they prevent the scalar originated from the real part of $\phi$ decay in these fields. If the other modes of this scalar field are suppressed then such a massive scalar state could be stable enough to be a kind of dark matter.

\begin{figure}[ht] 
\begin{center} 
\leavevmode 
\mbox{\epsfig{file=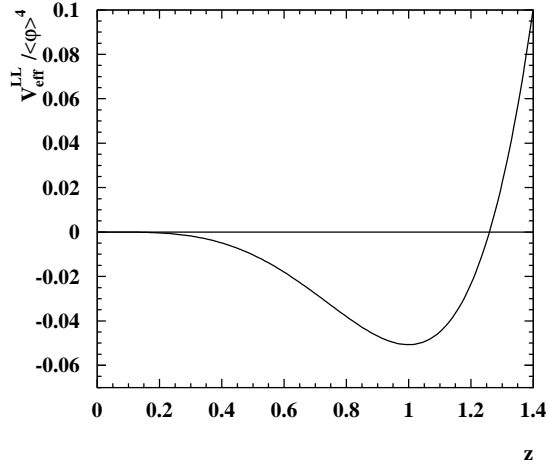,width=0.47
\textwidth,angle=0}}        
\end{center} 
\caption{$V_{eff}^{LL} /\langle\varphi\rangle^4 $ for the model when $\pi x_\alpha=0.010 $ and  $y=0.0649$. There is no significant change in the shape for the other values of $\pi x_\alpha$ and  $y$ in Table I.}
\label{figp} 
\end{figure}

Putting the renormalization scale as  the scalar field vacuum expectation value, i. e. $\mu= \langle\varphi\rangle$, and defining $z= \varphi/\langle\varphi\rangle$ the final expression for  $V_{eff}^{LL}$ is 

\begin{eqnarray}
V_{eff}^{LL}  \approx  {\pi^2} \langle\varphi\rangle^4 z^4 \left[ B\left( \ln z^2 - \frac{1}{2} \right) + C \ln^2z^2 + D \ln^3z^2 + E \ln^4 z^2  \right] 
\label{vefllz}
\end{eqnarray}
In Fig.\ref{figp} we plot  $V_{eff}^{LL} /\langle\varphi\rangle^4 $ for the model. 

The mass of the scalar field is 
\begin{eqnarray}
m_{\varphi} = 2\pi \sqrt{2(B+C)} \langle\varphi\rangle 
\label{ms}
\end{eqnarray}

We conclude this section saying that the effective potential in Eq. (\ref{vefllz}) breaks, besides the supposed scale invariance, also the U(1)$_{PQ}$ symmetry in Eq. (\ref{upq}). The condensation scale of the scalar field is then constrained by the limits on the axion mass. Taking, for example, the allowed upper bound permitted which is around $\langle\varphi\rangle\approx 10^{12}$ GeV we obtain $0.92\times 10^{12}$ GeV $\leq m_{\varphi}\leq$ $0.98\times  10^{12}$ GeV according to Table \ref{table1}. The CWM then  provides a dynamical mechanism to break the Peccei-Quinn symmetry.

\section{An application to an electroweak SU(3)$_L\otimes$U(1)$_X$ model}
\label{sec4}

An example of model that have a sector like Eq. (\ref{nuaxd}) is the one presenting a  SU(3)$_C\otimes$SU(3)$_L\otimes$U(1)$_X$  as in Ref. \cite{331ax1}, with a slight different discrete gauge symmetry $Z_{13}$. The electric charge operator defining the matter content is
\begin{equation}
{\cal Q}=T^3-\sqrt3 T^8+XI
\label{q}
\end{equation}
with $T^a$ the SU(3) generators. The multiplets are then:  
\begin{eqnarray}
\Psi_{aL}=(\nu_a,\,l_a,\,E_a)^T_L \sim({\bf1},{\bf3},0), 
\label{tsma}
\end{eqnarray}
$a=e,\mu,\tau$, representing the three left-handed leptonic triplets with the respective right-handed singlets 
\begin{eqnarray}
N^\prime_{aR} \sim({\bf1},{\bf1},0), \hspace{.7 cm}
l_{aR} \sim({\bf1},{\bf1},-1), \hspace{.7 cm}
E_{aR} \sim({\bf1},{\bf1},1);
\label{ssma}
\end{eqnarray}
left-handed quarks in the triplets are represented by 
\begin{eqnarray} 
Q_{mL}=(d_m,\, u_m,\, j_m)^T_L\sim({\bf3}, {\bf3}^{*},- 1/3), 
\hspace{.7 cm}
Q_{3L}=(u_3,\, d_3,\,J)^T_L\sim({\bf3}, {\bf 3}, 2/3), 
\label{tlqma} 
\end{eqnarray} 
$\;m=1,2$, with the respective right-handed singlets  as 
\begin{eqnarray} 
& & u_{\alpha R}\sim({\bf3},{\bf1},2/3), \hspace{.7 cm}
d_{\alpha R} \sim({\bf3},{\bf1},-1/3), \nonumber\\ 
& & J_{R}\sim({\bf3},{\bf1},5/3), \hspace{.7 cm}
j_{mR}\sim({\bf3},{\bf1},-4/3). 
\label{ssqma} 
\end{eqnarray} 
$J$ and $j_{m}$ are exotic quarks. The three triplets and a singlet of scalars needed to produce a consistent mass spectra are: 
\begin{eqnarray} 
& &\eta=(\eta^0,\,\eta^{-}_1,\,\eta^+_2)^T\sim({\bf1},{\bf3},0),\nonumber\\ 
& &\rho=(\rho^+,\,\rho^0,\,\rho^{++})^T\sim({\bf1},{\bf3},1), 
\nonumber\\ 
& &\chi=(\chi^-,\,\chi^{--},\,\chi^0)^T\sim({\bf1},{\bf3},-1) 
\nonumber\\ 
& &\phi\sim({\bf1},{\bf1},0).
\label{tsema} 
\end{eqnarray} 
This matter content, without the scalar singlet $\phi$, was first proposed  in Ref. \cite{331pt}. A discrete $Z_{13}$ symmetry with a local origin is supposed  over the Lagrangian. Their charges under this symmetry factor are shown in Table \ref{table2}

\begin{table}
\begin{tabular}{||c|c|c|c|c|c|c|c|c|c|c|c|c|c||}\hline
$\Psi_L$ & $\phi$ & $d_{\alpha R}$ & $\chi$ & $J_R$ & $\rho$ & $Q_{3L}$ & $\eta $ & $u_{\alpha R}$ & $E_R$  & $j_{mR}$  & $l_R$  & $Q_{iL}$ & $N^\prime_R$ \\
\hline
0 & 1 & 1 & 2 & 3 & 4 & 5 & 6 & -1 & -2 & -3 & -4 & -5 & -6 \\ \hline
\end{tabular}
\caption{$Z_{13}$ charges for the multiplets.}
\label{table2}
\end{table}

These charges permit to construct a Yukawa potential containing all needed renormalizable interactions among fermions and scalars to generate a consistent mass spectra. The Yukawa terms involving the leptons are 
\begin{eqnarray}
-{\cal L}^l_Y & = &  G^l_{ab} \overline{(\Psi) }_{aL}l_{bR}\rho
+ G^{^E}_{ab} \overline{(\Psi) }_{aL}E_{bR}\chi \nonumber \\ 
& + & G^N_{ab} \overline{(\Psi)}_{aL}N^\prime_{bR}\,\eta + \frac{\lambda^{\prime}_{_{ \alpha\beta}}}{2}\phi^* \overline{(N^\prime_{_{ \alpha R}})^c}N^\prime_{_{ \beta R}} + H.c,
\label{yulep}
\end{eqnarray}
with $G^N_{ab}$, $G^l_{ab}$ and $G^{^E}_{ab}$ arbitrary complex matrices. We see that the last two terms in Eq. (\ref{yulep}) are the same as in Eqs. (\ref{nuaxd}) and (\ref{nuNdirac}) which are required for the see-saw mechanism. The quarks terms can be written in a straightforward manner but, since they are not relevant for us here, we will omit them.

The renormalizable, $Z_{13}$ and scale invariant scalar potential is 
\begin{eqnarray}
V_{\rm 331} & = & \lambda_1(\eta^\dagger \eta)^2+ \lambda_2(\rho^\dagger \rho)^2 + \lambda_3(\chi^\dagger \chi)^2 + \eta^\dagger\eta\left[\lambda_4\rho^\dagger\rho + \lambda_5\chi^\dagger\chi\right] + \lambda_6 \left(\rho^\dagger\rho\right)\left(\chi^\dagger\chi\right) 
\nonumber \\
& + & \lambda_7\left(\rho^\dagger\eta\right)\left(\eta^\dagger\rho\right) + \lambda_8\left(\chi^\dagger\eta\right)\left(\eta^\dagger\chi\right) + \lambda_9\left(\rho^\dagger\chi\right)\left(\chi^\dagger\rho\right) + \lambda_{10}\left[\epsilon^{ijk} \phi\eta_i\rho_j\chi_k + \mbox{H. c.}\right] 
\nonumber \\ 
&+& \phi^*\phi\left[\lambda_{ \phi\eta} \eta^\dagger\eta + \lambda_{ \phi\rho} \rho^\dagger
\rho + \lambda_{ \phi\chi} \chi^\dagger\chi \right] + \lambda_\phi (\phi^*\phi)^2 
\label{pe}
\end{eqnarray}

One can see that Eqs. (\ref{yulep}) and (\ref{pe}) automatically contain a kind of global U(1)$_{PQ}$ symmetry for solving the strong CP problem. The transformations are 
\begin{eqnarray}
& & \nu_L\rightarrow e^{i \delta}\nu_L, \; \; \; N^{\prime}_{aR}\rightarrow e^{-i\delta}N_{aR}, \; \; \; l \rightarrow  e^{i\delta\gamma_5}l,\nonumber \\
& &   u_a \rightarrow e^{-i\delta\gamma_5}u_a, \;\; \; d_a \rightarrow e^{i\delta\gamma_5}d_a,\;\;\; 
j_m \rightarrow e^{i\delta\gamma_5}j_m, \nonumber \\
& &  E_a\rightarrow e^{-i\delta\gamma_5}E_a, \;\; \; 
J\rightarrow e^{-i\delta\gamma_5}J,
\label{pqfma}
\end{eqnarray}
The scalars with nontrivial U(1)$_{PQ}$ transform as 
\begin{eqnarray}
& &  \eta^{0}\rightarrow   e^{ 2i\delta}\eta^0, \;\; \; \eta^{+}_2\rightarrow  e^{2i\delta}\eta^+_2, \;\; \; \rho^{0}\rightarrow   e^{-2i\delta}\rho^0, \nonumber \\
& &  \chi^{-}\rightarrow  e^{2i\delta}\chi^-, \;\;\;   \chi^{0}\rightarrow   e^{2i\delta} \chi^0, \;\;\; \phi\rightarrow   e^{-2i\delta}\phi. 
\label{pqema}
\end{eqnarray}
As we see, the scalar singlet $\phi$ is charged under this U(1)$_{PQ}$. Also, the last term in Eq. (\ref{yulep}) does not conserve the lepton number. Even if a lepton number was assigned to $\phi$ in order to make that term invariant, the explicit break would be transferred to the $\phi\eta\rho\chi$ interaction in Eq. (\ref{pe}). 

The model then has a sector like that in Eq. (\ref{nuax}) and assuming that the condensation of $\phi$ must occur in the appropriated scale inside the window permitted for the invisible axion, it is well justified, at first sight, the fact that  $\phi$ and the $N_{aR}$ form a sector decoupled from the rest of the fields. Even though there are the couplings $\phi^\dagger\phi\eta^\dagger\eta$, $\phi^\dagger\phi\rho^\dagger\rho$, $\phi^\dagger\phi\chi^\dagger\chi$ and $\phi\eta\rho\chi$ their respective coupling constants $\lambda_{\phi\eta}$, $\lambda_{\phi\rho}$, $\lambda_{\phi\chi}$, $\lambda_{10}$, are suppressed by factors of the VEV's $\langle\eta\rangle$, $\langle\rho\rangle$ and  $\langle\chi\rangle$ divided by $\langle\phi\rangle$ as we are going to see in the following. 

Assuming, then, that the scales of the scalar triplets responsible for the break of ${\textrm{SU(3)}}_L\otimes{\textrm{U(1)}}_N$ down to the electromagnetic factor are in the TeV-GeV range, which is the interval expected for arising some new physics, the separation of the scales permits the calculation of the effective potential for $\phi$ independently of the condensation remaining scalars. Therefore, the potential below $\langle\phi\rangle$ is 
\begin{eqnarray}
V_{\rm 331} & \simeq & V_H(\eta, \rho, \chi) + \frac{\lambda_{10}}{\sqrt2}\left[\epsilon^{ijk} \varphi e^{i\frac{a(x)}{\langle\phi\rangle}} \eta_i\rho_j\chi_k + \mbox{H. c.}\right] 
\nonumber \\ 
&+& \frac{1}{2}\varphi^2\left[\lambda_{ \phi\eta} \eta^\dagger\eta + \lambda_{ \phi\rho} \rho^\dagger
\rho + \lambda_{ \phi\chi} \chi^\dagger\chi \right] + V_{eff}^{LL}(\varphi)  
\label{pef}
\end{eqnarray}
with $V_H(\eta, \rho, \chi)$ collecting all hermitian terms involving only the triplets and $V_{eff}^{LL}(\varphi)$ like Eq. (\ref{vefllz}). Observe that the bilinears bringing the quadratic mass terms to the triplets have their origin in the interaction with $\varphi$ through its condensation. Allowing now the SSB in the usual way by giving VEV to the remaining neutral scalars also, by means of the potential minimization condition $\frac{d\,V_{eff}^{LL}}{d\, \xi}{\Big{\vert}}_{_{\{\xi= \langle\xi\rangle\} }}= 0$, $\xi=\eta^0, \rho^0, \chi^0, \varphi$, the four constraint equations obtained are 
\begin{eqnarray}
&& \left[2\lambda_1v_\eta^2 +\lambda_4v^2_\rho+\lambda_5v^2_\chi+\lambda_{\phi\eta} 
v_\phi^2\right]v_\eta+
\lambda_{10}v_\rho v_\chi v_\phi = 0\,,
\nonumber \\
&& \left[2\lambda_2v_\rho^2+\lambda_4 v_\eta^2+\lambda_6 v_\chi^2+\lambda_{\phi\rho} 
v_\phi^2\right]v_\rho+
\lambda_{10}v_\eta v_\chi v_\phi = 0\,,
\nonumber \\ 
&& \left[2\lambda_3v_\chi^2+\lambda_5  v_\eta^2+
\lambda_6  v_\rho^2+\lambda_{\phi\chi}  v_\phi^2\right]v_\chi+\lambda_{10} v_\eta v_\rho v_\phi = 0\,, \nonumber \\&& \left[\lambda_{\phi\eta} v_\eta^2 + 
\lambda_{\phi\rho} v_\rho^2+ \lambda_{\phi\chi} v_\chi^2\right]v_\phi+\lambda_{10} v_\eta v_\rho v_\chi = 0\,,
\label{ce}
\end{eqnarray}
where we have taken, $\langle \varphi \rangle = v_{\phi}$, 
$\langle \chi^0 \rangle = \frac{1}{\sqrt2}v_{\chi}$, 
$\langle \eta^0 \rangle = \frac{1}{\sqrt2}v_{\eta}$ and 
$\langle \rho^0 \rangle = \frac{1}{\sqrt2}v_{\rho}$. These constrains are the same obtained recently in Ref. \cite{potef331} in a model with the same scalar sector without using the improved approximation. The fact that Eqs. (\ref{ce}) do not contain any term due to $V_{eff}^{LL}(\varphi)$ is because this last one is already at the minimum in the $\langle \varphi \rangle$ direction in the field space. Thus, the conclusions over the VEV's are the same as in \cite{potef331}. Once $v_{\phi}$ is different from zero according to CWM, the constraint equations above can be satisfied only by two possible configurations. The first one is the trivial one, $v_{\chi}=v_{\eta}=v_{\rho}=0$, excluded phenomenologically. The second one is when all the VEV's are non-zero and this is a full consistent solution. This means that in this scheme the VEV $\langle \chi^0 \rangle$  cannot break ${\textrm{SU(3)}}_L\otimes{\textrm{U(1)}}_N$  leaving behind $\textrm{SU(2)}_L\otimes{\textrm{U(1)}}_Y$ intact. 

The couplings of $\phi$ with the other scalars are then determined from Eqs. (\ref{ce}). They are 
\begin{eqnarray}
\lambda_{\phi\eta} &=& \frac{1}{v_\eta^2 v_\phi^2} \left[ -\lambda_1 v_\eta^4 + \lambda_2 v_\rho^4 + \lambda_3 v_\chi^4 + \lambda_6 v_\rho^2 v_\chi^2 \right],
\nonumber \\
\lambda_{\phi\rho} &=& \frac{1}{v_\rho^2 v_\phi^2} \left[ \lambda_1 v_\eta^4 - \lambda_2 v_\rho^4 + \lambda_3 v_\chi^4 + \lambda_5 v_\eta^2 v_\chi^2 \right],
\nonumber \\
\lambda_{\phi\chi} &=&  \frac{1}{v_\chi^2 v_\phi^2} \left[ \lambda_1 v_\eta^4 + \lambda_2 v_\rho^4 - \lambda_3 v_\chi^4 + \lambda_4 v_\eta^2 v_\rho^2 \right],
\nonumber \\
\lambda_{10} &=& - \frac{1}{v_\eta v_\rho v_\chi v_\phi} \left[ v_\eta^2 (\lambda_1 v_\eta^2 + \lambda_4 v_\rho^2) + v_\rho^2(\lambda_2 v_\rho^2 + \lambda_6 v_\chi^2) + v_\chi^2 (\lambda_3 v_\chi^2 + \lambda_5 v_\eta^2) \right]
\label{coup}
\end{eqnarray}
As we have anticipated, these coupling constants are suppressed by factors of the triplet scale divided by $v_\phi$. Thus, the couplings of $\phi$ can be safely ignored in the calculation  of $V_{eff}^{LL}(\varphi)$ for this model and the results in Sec. \ref{sec3} can be well applied. 

The see-saw mechanism occurs in the standard way. After the neutral components in the triplets get VEV the mass terms for the neutrinos are  
\begin{eqnarray}
{\cal{L}}^{mass}_{\nu} &=&   G^N_{_{\alpha\beta}} \frac{v_\eta}{\sqrt2} \overline{\nu_{_{\alpha L} }}N^\prime_{_{ \beta R}} +  \frac{\lambda^{\prime} _{_{ \alpha\beta}}}{2}\frac{v_\phi}{\sqrt2}\overline{( N^\prime_{_{ \alpha R}})^c} N^\prime_{_{ \beta R}}+H.c.
\nonumber\\
 \nonumber\\
 &=&  \frac{G^N_{_{\alpha\beta}}}{2} \frac{v_\eta}{\sqrt2}[ \overline{\nu_{_{\alpha L} }} N^\prime_{_{ \beta R}} + \overline{( N^\prime_{_{\beta R} })^c}( \nu_{_{ \alpha L}})^c ]  +    \frac{\lambda^{\prime} _{_{ \alpha\beta}}}{2} \frac{v_\phi}{\sqrt2}\overline{( N^\prime_{_{ \alpha R}})^c} N^\prime_{_{ \beta R}}+H.c.
\nonumber\\
 \nonumber\\
&=&  \frac{1}{2}  \left( \begin{array}{cc}
 \overline{ \nu_{_{ L} }} & \overline{( N_{_{ R} })^c}
\end{array} \right)
\left( \begin{array}{cc}
0 &  {\cal{M}}_D \\
{\cal{M}}_D^T  &  {\cal{M}}_M \end{array} \right)
\left( \begin{array}{c}
 (\nu_{_{ L}})^c  \\
 N_{_{ R}}
\end{array} \right) + H.c.
\label{nusmass}
\end{eqnarray} 
where $ \nu_{_{ L}}\equiv ( \nu_{_{e L}}\;\; \nu_{_{\mu L}}\;\; \nu_{_{\tau L}})^T$ and $ N_{_{ L}}\equiv ( N_{_{e L}}\;\; N_{_{\mu L}}\;\; N_{_{\tau L}})^T$; and the Dirac and Majorana mass matrices are defined as 
\begin{equation}
[{\cal{M}}_D]_{\alpha\beta} = G^N_{\alpha\sigma}[U_N]_{\sigma\beta}  \frac{v_\eta}{\sqrt2}
\label{numassd}
\end{equation}
\begin{equation}
[{\cal{M}}_M]_{\alpha\beta} = [U^T_N \lambda^{\prime}U_N]_{{\alpha\beta}} \frac{v_\phi}{\sqrt2}   
\label{numassm}
\end{equation}
Just for an estimate, supposing ${\cal{M}}_D$ and ${\cal{M}}_M$ being diagonal, with entries of the same order, then the active neutrinos would have mass
\begin{equation}
m_{\nu_{_{\alpha L}}}\simeq G^N_{\alpha\sigma}[U_N]_{\sigma\alpha}  \frac{v^2_\eta}{\sqrt2 v_\phi} 
\label{mnuma}
\end{equation} 
If $v_\phi \approx 10^{12}$ GeV and $v_\eta\approx 10^2$ GeV then we could have $G^N_{\alpha\sigma}[U_N]_{\sigma\alpha}\approx 10^{-2}$ to furnish $m_{\nu_{_{\alpha L}}}\simeq 0.1$ eV, and so it seems that no extreme fine tuning over the $G^N_{\alpha\sigma}$ is required to produce small masses.

\section{Final discussion}
\label{sec:disc}

We have seen that the generic interaction sector composed by a scalar field coupled with Majorana fermions, both singlets of the electroweak group, with the tree level potential $V=\lambda(\phi^\dagger \phi)^2$ generates an effective potential breaking  scale invariance and the U$_{PQ}$ symmetry. The calculation was performed at one loop with the summation of all leading-logarithms and we have obtained a stable effective potential inside the pertubative domain. Although a more satisfactory answer concerning the stability of the potential would be obtained only considering a calculation beyond one loop approximation also taking the subsequent-to-leading logarithmic corrections, we believe that the result will be maintained when higher order corrections were taken into account. This has been shown to be true for the case of the Higgs doublet in the SM \cite{elias05}. But, of course, a complete calculation must be done in order to settle the question in the model treated here. It will be done elsewhere.

In the numerical example we took, we have seen that there is a region in the  space parameter where the massive scalar composed mainly by the real part of $\phi$ cannot decay in two  Majorana fermions. So, it could be that such a scalar has some relevance as some sort of dark matter. 

A specific model based on SU(3)$_L\otimes$U(1)$_X$ symmetry implemented  with a supposed gauge discrete symmetry, to stabilize the invisible axion solution to the strong CP problem, realizes the whole idea. The complex scalar singlet in this model was shown to have very suppressed couplings with all other scalars as we see in Eq. (\ref{coup}). This is due to the large VEV $v_\phi$ and the initial scale invariance so that only the singlet Majorana fermions have leader contribution for the effective potential. Furthermore, the CP odd mass spectra of the scalars has the axion and one massive state which could be expected to belong to the scale $v_\phi$. However, because Eq. (\ref{coup}) this massive scalar in fact belongs to the $v_\chi$ scale. The only fields with mass proportional to $v_\phi$ are the CP even scalar composed mainly by the real part of $\phi$ and the singlet Majorana fermions.

Finally, we call the attention that besides realizing the see-saw mechanism as a result of the dynamical condensation in the sector that we have discussed, the model in Sec. \ref{sec4} shows electric charge quantization as a result of the classical constraints in the Yukawa Lagrangian and the quantum anomalies cancellation conditions \cite{mohaq}, \cite{piresq1}.  In other words, it is possible to fix the normalization, $q_X$, of the factor U(1)$_X$ when we leave all the respective $X$ charges of the multiplets in Eqs. (\ref{tsma}), (\ref{ssma}), (\ref{tlqma}), (\ref{ssqma}) and (\ref{tsema}).  Writing  the electric charge operator now as 
\begin{equation}
{\cal Q}=T^3-\sqrt3 T^8+q_{X} XI
\label{qq}
\end{equation}
and using the fact that it annihilates the vacuum,
\begin{equation}
{\cal Q}\langle \xi \rangle= 0
\label{qarb0}
\end{equation}
with $\xi= \eta,\rho,\chi,\phi$, we get that the scalar multiplets must be such that 
\begin{equation}
q_X = \frac{1}{X_\rho},\,\,\,\,\, X_\eta=X_\phi= 0,\,\,\,\,\,  X_\rho=-X_\chi .
\label{xesc}
\end{equation} 
With the quarks the Yukawa Lagrangian is 
\begin{eqnarray}
-{\cal L}^q_Y &=& 
\overline{Q}_{iL} (F_{i\alpha }u_{\alpha
R}\rho ^{*}+\widetilde{F}_{i\alpha } d_{\alpha R}\eta ^{*}) 
+ \overline{Q}_{3L} ( G_{\alpha }
u_{\alpha R}\eta +
\widetilde{G}_{\alpha } d_{\alpha R}\rho) \nonumber \\ 
&+& \lambda^{^J}  \overline{Q}_{3L}J_{1R} \chi+
\lambda^j _{im} \overline{Q}_{iL}j_{mR}\chi ^{*}+H.c.,
\label{yuqma}
\end{eqnarray} 
with $F_{i\alpha }$, $\widetilde{F}_{i\alpha }$, $G_{\alpha }$, $\widetilde{G}_{\alpha }$,  $\lambda^{^J}$ and $\lambda^j _{im}$ arbitrary matrices. The above interactions in addition with those in Eq. (\ref{yulep}) and the relations in (\ref{xesc}) determine all the U(1)$_X$ charges in function of two of them, say $X_\rho$ and $X_Q$. Now to cancel the $\left[ {{\textrm{SU(3)}}_L}\right]^2{\textrm{U(1)}}_X$ anomaly we get the missing relation $X_{Q} =- \frac{1}{3} X_\rho$ establishing, thus, electric charge quantization.

The author is supported by FAPESP under the process 05/52006-6.

\end{document}